\documentclass[11pt]{JHEP}
\usepackage{amssymb,amsfonts}

\renewcommand{\thesection}{\arabic{section}}

\preprint{August 11, 2013 \\ V2. November 26, 2013. \\ To appear in JHEP }
\title{Non--Abelian tensor hierarchy in (1,0) D=6 superspace}

\author{ Igor A. Bandos $^{\ast\dagger}$ \\  {\small\it $^{\ast}$ Department of Theoretical Physics,
University of the Basque Country,} ~\\ {\small\it UPV/EHU, P.O. Box 644, 48080 Bilbao, Spain}
{\small\it and} ~\\ {\small\it $^{\dagger}$ IKERBASQUE, Basque Foundation for Science, 48011, Bilbao,
Spain} }

\vspace{2cm}

\abstract{We present a set of constraints on superfield strengths of the non-Abelian p-form potentials
in D=6 (1,0) superspace which reproduces, as their selfconsistency conditions, the equations of motion
of the recently proposed (1,0) superconformal theory. These  include the anti-self-duality conditions
for the field strength of the non-Abelian 2-form potential, duality between field strengths of the
non--Abelian vectors and 3-forms as well as of the non-Abelian four forms  and  scalar fields.}

\keywords{Supersymmetry, superspace, non--Abelian tensor fields, duality and self-duality,  multiple
p-branes}

\begin{document}

\section{Introduction}

Recently, motivated by the search for a description of multiple M5-brane system
\cite{Witten:1995zh,Lambert:2010wm,Lambert:2010iw,Douglas:2010iu,Singh:2011id,Lambert:2011gb,Lambert:2012qy,
Ho:2011ni,Huang:2012tu,Chu:2012um,Chu:2012rk,Chu:2013hja,Bonetti:2012fn,Bonetti:2012st,Singh:2012qr,
Kim:2012tr}, the authors of \cite{Samtleben:2011fj,Samtleben:2012mi,Samtleben:2012fb} have constructed
a class of new (1,0) superconformal models  describing a hierarchy of non--Abelian scalar, vector and
tensor fields and their supersymmetric partners. The action for the bosonic sectors of these theories
have been constructed in a very recent \cite{BSS2013} using the PST (Pasti--Sorokin-Tonin ) approach
\cite{Pasti:1996vs}.

In this paper we propose the set of constraints on the super-(p+1)--form field strengths of
non--Abelian super-p-form potentials on (1,0) D=6  superspace which restrict the field content of these
super-p-forms to the fields of the non-Abelian tensorial hierarchy of \cite{Samtleben:2011fj}. We show
that these constraints reproduce the dynamical equations of the (1,0) superconformal model as their
selfconsistency conditions. The set of these equations includes supersymmetrizations of the
anti--self--duality condition for the 3-form field strength of non-Abelian 2-form (antisymmetric
tensor) potential, as well as non-Abelian vectors---3-forms and scalars---4-forms duality relations.

Although the same equations were obtained in \cite{Samtleben:2011fj} from closure of the algebra of
supersymmetry transformations on the spacetime fields, the superfield formulation of tensorial
hierarchy may be useful as it clarifies the structure of the theory. In particular, it provides a basis
for the search \cite{BSS2013+} for supersymmetric generalization of the action of \cite{BSS2013} and,
hopefully, can provide an insight in looking for a (2,0) superconformal theory  by superspace methods.

\section{Non-Abelian $p$--forms in six dimensions} \setcounter{equation}{0}

The (1,0) superconformal 6d field theories of \cite{Samtleben:2011fj,Samtleben:2012mi,Samtleben:2012fb}
describe a hierarchy of non--Abelian scalar, vector and tensor fields $( Y^{ij\,r},\phi^I,A_\mu^r,
B_{\mu\nu}^I, C_{\mu\nu\rho\,r},C_{\mu\nu\rho\lambda\, A})$ and their supersymmetric partners.
The upper indices $r,s, t=1,..., n_V$ enumerate vector multiplets. Generically, the gauge group is not semi-simple: it may have the structure of direct product and contain Abelian factors. The indices $I,J,K=1,..., n_T$ enumerate the tensor multiplets. The index $A$ is used to enumerate 4-forms, while the three forms are enumerated by the lower $r,s,t$ indices.

As in
\cite{BSS2013} we will use the differential form notation in which the bosonic field content can be
described by spacetime differential $p$--forms (0-form corresponds to a scalar) as
$(Y^{ij\,r},\phi^I,A^r_1,B_2^I,C_{3\,r},C_{4A})$. This is especially convenient for the description of
tensorial hierarchy in superspace, because the differential form equations which do not involve the
Hodge star operator can be manipulated without referring on what is the base manifold (supermanifold).

So let us define the generalized field strengths \begin{eqnarray}\label{cF:=} {\cal F}^r&=&{\cal
F}_2^r=dA^r+{1\over 2}f_{rs}{}^t A^s\wedge A^t + g_I^rB^I_2 , \qquad \\ \label{cH3:=}
 {\cal H}_3^I&=& dB^I_2+d_{st}^I A^s\wedge dA^t +... + g^{Ir}C_{3r}\; , \qquad \\
\label{cH4:=}  {\cal
H}_{4r}&=& dC_{3r}+...+ k_r^AC_{4A} \;  \quad
\end{eqnarray}
of the non-Abelian Yang-Mills, two form and three form
potentials $A^r:=A_1^r$, $B_2^I$ and $C_{3r}$ by stating that they obey the Bianchi identities
\begin{eqnarray}\label{DcF=} &&
I_3^r:= D{\cal F}^r- g_I^r {\cal H}_3^I =0 \; , \qquad
\\ \label{DcHI=}
&& I_4^I:= D{\cal
H}_3^I-
 d_{st}^I {\cal F}^s\wedge {\cal F}^t-g^{Ir} {\cal H}_{4r} =0
\; , \qquad \\ \label{DcH4=} && I_{5r}:= D{\cal H}_{4r} + 2{\cal H}_3^I \wedge {\cal F}^sd_{Isr}- k_r^{A}{\cal
H}_{5A}=0\; . \qquad \end{eqnarray}
Here $f_{rs}^t=f_{[rs]}^t$, $d_{st}^I=d_{(st)}^I$ and $g^r_I$ are
constant and covariantly constant tensors. The list of their properties can be found in  Appendix A  as well as in the original paper  \cite{Samtleben:2011fj}.
The meaning of $f_{rs}^t$ is the structure constant
of the gauge algebra, while $d_{st}^I$ defines the nonlinear gauge field contribution to the 3--form field
strength of the two form potential; $g_I^r$, $g^{Ir}$ and $ k_r^{A}$ are St\"uckelberg  couplings.
We restrict ourselves by the case of tensorial
hierarchy which allows for existence of an action, this is to say we assume the existence of the (not
positively definite, Lorentz-type) metric $\eta_{IJ}=\eta_{(IJ)}$, so that $g_I^r=\eta_{IJ}g^{Jr}$.

The knowledge of Bianchi identities (\ref{DcF=})--(\ref{DcH4=}) is  completely sufficient for the
discussion below; we will not need the complete explicit expressions (\ref{cF:=})--(\ref{cH4:=}) for the field
strengths in terms of the non-Abelian p-form potentials (see \cite{Samtleben:2011fj} and \cite{BSS2013}
where the equations from \cite{Samtleben:2011fj} are written in differential form notations). What we
need is rather the explicit form of the covariant derivatives $D$ which are used in (\ref{DcF=}),
(\ref{DcHI=}), (\ref{DcH4=}), \begin{eqnarray}\label{DFr=d+} && D{\cal F}^r:=d{\cal F}^r + {\cal
F}^t\wedge A^s X_{st}{}^r  =d{\cal F}^r - {\cal F}^t\wedge A^s f_{st}{}^r + {\cal F}^t\wedge A^s
d_{st}^Ig_I{}^r \; , \qquad \\ \label{DcHI=d+} && D{\cal H}^I_3:=d{\cal H}^I_3 + {\cal H}^J_3\wedge A^s
X_{sJ}{}^I =d{\cal H}^I_3 + 2{\cal H}^J_3\wedge A^s d_{st}^Ig_J{}^t  - 2{\cal H}^J_3\wedge A^r
g^{Is}d_{Jsr} \; , \qquad\\ &&D{\mathcal H_{4\,r}}:=d{\mathcal H_{4\,r}}- \mathcal H_{4\,t}\wedge A^s
X_{sr}{}^t\,.\label{DH4} \end{eqnarray}
In our notation the exterior derivative acts from the right, so that, {\it e.g.}, $d({\cal F}^r\wedge  {\cal H}_3^I)= {\cal F}^r\wedge  d{\cal H}_3^I- d{\cal F}^r\wedge  {\cal H}_3^I$.

Below we will consider the potential defined on the flat (1,0) D=6 superspace which we are going to
describe now.

\section{Tensor hierarchy in superspace} \setcounter{equation}{0}

\subsection{6d (1,0) superspace}

The structure equations of  flat 6D ${\cal N}=(1,0)$ superspace $\Sigma^{(6|8)}$ are
\begin{eqnarray}
\label{DEa=} dE^a= - i E^{\alpha i}\wedge E^{\beta j} \gamma^a_{\alpha\beta}\epsilon_{ij}\; , \qquad
dE^{\alpha i}=0\; .
\end{eqnarray}
Here  $E^a$ and  $E^{\alpha i}$ denote  6 bosonic
and 8 fermionic supervielbein 1--forms of $\Sigma^{(6|8)}$, $\epsilon_{ij}=- \epsilon_{ji}$ is
normalized by $\epsilon^{12}=1=-\epsilon_{12}$ and
\begin{equation}\label{gamma=etg}
\gamma^a_{\alpha\beta}=-\gamma^a_{\beta\alpha}= {1\over
2}\epsilon_{\alpha\beta\gamma\delta}\tilde{\gamma}^{a\gamma\delta}
\end{equation} are $SO(1,5)$
Klebsh-Gordan coefficients (generalized Pauli matrices) which obey
\begin{eqnarray}
 \label{6Dgammas}
(\gamma^{a}\tilde{\gamma}{}^{b }+ \gamma^{b}\tilde{\gamma}{}^{a})_{\alpha}{}^{\beta} =
2\eta^{ab}\delta_{\alpha}{}^{\beta}\; ,  \qquad \eta^{ab}= diag (+,-,-,-,-,-)\; , \qquad \nonumber \\
\gamma^a_{\alpha\beta}\tilde{\gamma}_a^{\gamma\delta}= -4 \delta_{[\alpha}{}^{\gamma} \delta_{\beta
]}{}^{\delta} \; , \qquad \gamma^a_{\alpha\beta} {\gamma}_{a \gamma\delta}= -2
\epsilon_{\alpha\beta\gamma\delta} \; , \qquad \\ \gamma^{abcdef}{}_{\alpha}{}^{\beta}= -
\epsilon^{abcdef}\delta_{\alpha}{}^{\beta}  \; . \qquad \nonumber \end{eqnarray} Notice that
 \begin{eqnarray}\label{6Dg3=sd}
\gamma^{abc}{}_{\alpha\beta}=\gamma^{abc}{}_{(\alpha\beta )}= + {1\over
3!}\epsilon^{abcdef}\gamma_{def\;\alpha\beta} \; , \qquad \tilde{\gamma}^{abc
\alpha\beta}=\tilde{\gamma}^{abc (\alpha\beta )}=  - {1\over
3!}\epsilon^{abcdef}\tilde{\gamma}_{def}^{\alpha\beta} \;  \qquad \end{eqnarray} are self-dual and
anti--self--dual, respectively, with respect to their antisymmetrized vector indices,  and provide the complete basis for the symmetric $4\times 4$ matrices with, respectively, two lower case and two upper-case 4-valued spinor  indices $\alpha,\beta,... =1,...,4$.  The other useful relations can be found in the Appendix B.

The structure equations can be easily solved by \begin{eqnarray}\label{Ea=} E^a= dx^a- i d\theta^{i}
\gamma^a\theta_i\; , \qquad E^{\alpha i}=d\theta^{\alpha i}\; , \end{eqnarray} where
$Z^{M}=(x^m,\theta^{{\alpha} i})$ are local coordinates of $\Sigma^{(6|8)}$  and $\theta_i^{\beta}:=
\epsilon_{ij}\theta^{\beta j} $ so that $ d\theta^{i} \gamma^a\theta_i= d\theta^{\alpha i} \,
\theta^{\beta j} \epsilon_{ij}\gamma^a_{\alpha\beta}$.

\subsection{Constraints for the superspace field strengths}

When our field strengths, and corresponding non-Abelian $p$--form potentials, are differential forms on
superspace $\Sigma^{(6|8)}$, they can be decomposed  on the basis of wedge products of the
supervielbein forms (\ref{Ea=}),  $E^{{\cal A}}=(E^{a}, E^{\alpha i}):= dZ^M E^{{\cal A}}(Z)$,
\begin{eqnarray}
 \label{cF=SSP} {\cal F}^r=
{1\over 2} E^{{\cal B}}\wedge E^{{\cal A}} {\cal F}^r_{{\cal A}{\cal B}}(Z) \; ,  \qquad {\cal H}_3^I=
{1\over 3!}E^{{\cal C}}\wedge E^{{\cal B}}\wedge E^{{\cal A}} {\cal H}_{{\cal A}{\cal B}{\cal
C}}^I(Z)\; , \qquad \nonumber
 \\ {\cal H}_{4r}= {1\over 4!}E^{{\cal D}}\wedge E^{{\cal C}}\wedge E^{{\cal B}}\wedge E^{{\cal A}}
 {\cal H}_{{\cal A}{\cal B}{\cal C}{\cal D}\; r}(Z)\; . \qquad
\end{eqnarray}
To restrict the huge field content of the generic super--p--form potentials to the fields of the (1,0)
superconformal theory of \cite{Samtleben:2011fj}, we  impose  the following set of constraints
\begin{eqnarray}\label{cF=ssp} {\cal F}^r &=& iE^b\wedge E^{\alpha i} (\gamma_bW_i^r)_\alpha + {1\over
2} E^b\wedge E^a {\cal F}^r_{ab}\; , \qquad \\ \label{cH3=ssp} {\cal H}_3^I &=& {i\over 2} E^b\wedge
E^{\alpha i} \wedge E^{\beta j}  \gamma_{b\alpha\beta} \epsilon_{ij} \Phi^I +{i\over 2} E^b\wedge E^a
\wedge E^{\alpha i} (\gamma_{ab}\Psi^I_i)_\alpha +  {1\over 3!} E^c\wedge E^b\wedge E^a {\cal
H}^I_{abc}\; , \quad\\ \label{cH4=ssp} {\cal H}_{4r} &=&
 -{i\over 3!}E^c \wedge E^b\wedge E^a \wedge E^{\alpha i}{\gamma}_{abc\alpha \beta}W_i^{\beta
 s}d_{Isr}\Phi^I  + {1\over 4!} E^d\wedge  E^c\wedge E^b\wedge E^a {\cal H}_{abcd\, r}
 \; . \qquad
\end{eqnarray}
Here $W_i^{\alpha \; r}$ and $\Psi^I_{\alpha\; i}$ are fermionic spinorial superfields, while
$\Phi^I$, ${\cal F}^r_{ab}={\cal F}^r_{[ab]}$,  ${\cal
H}^I_{abc}={\cal
H}^I_{[abc]}$ and ${\cal H}_{abcd\, r}={\cal H}_{[abcd]\, r}$  are bosonic scalar and antisymmetric tensor superfields which at this stage can be considered unrestricted. The leading components of these superfields will give rise to the physical fermionic fields of vector and tensor multiplets
($\chi^{\alpha \; r}(x)= W_i^{\alpha \; r}\vert_{\theta =0}$ and $\psi^I_{\alpha\; i}(x)=\Psi^I_{\alpha\; i}\vert_{\theta =0}$),  to the  scalar field of the tensor multiplet ($\phi^I(x)\propto \Phi^I\vert_{\theta =0}$), and to the field strengths of the vector gauge fields and of the higher form potentials of the tensorial  hierarchy.

Actually, the above expressions for the superform field strengths  collect the
independent constraints together with some of their consequences. In particular, the true constraints
on the supersymmetric  Yang--Mills (SYM) field  strength are ${\cal F}^r_{\alpha i\; \beta j}=0$, which imply
\begin{eqnarray}\label{DfDf=Db}
 \{ {\cal D}_{\alpha i}, {\cal D}_{\beta j} \} = 2 i \epsilon_{ij}\gamma^a_{\alpha\beta}{\cal D}_{a}\;
 .
\end{eqnarray} Then the expression for the field strength 2-form reflecting this constraint is ${\cal
F}^r= iE^b\wedge E^{\alpha i} {\cal F}^r_{\alpha i\; b}  + {1\over 2} E^b\wedge E^a {\cal F}^r_{ab}$.
The fact that ${\cal F}^r_{\alpha i\; b}= i (\gamma_bW_i^r)_\alpha$, as it is read from (\ref{cF=ssp}),
follows as the solution of the selfconsistency conditions given by Bianchi identities.

As another example, the true constraints in Eq. (\ref{cH4=ssp}) are  ${\cal H}_{ \alpha i\; \beta
j\;{\cal C}{\cal D}\;r}=0$, while the expression for  ${\cal H}_{ \alpha i\; bcd \;r}$, presented in
the first term of Eq. (\ref{cH4=ssp}), is obtained by studying their selfconsistency conditions given
by the Bianchi identities\footnote{ Actually,  one can guess the possible structure of the first
nonvanishing term in Eq. (\ref{cH4=ssp}), write it with an arbitrary coefficient $il$, and fix
$l=-{1\over 6}$ by studying  the Bianchi identities.}.

\subsection{Equations of motion from consistency of the superspace constraints}

Studying the Bianchi identities (\ref{DcF=}) with the constraints (\ref{cF=ssp}) and (\ref{cH3=ssp}) we
find the structure of covariant derivatives of fermionic  superfield of the  SYM  sector,
\begin{eqnarray}\label{DfW=} {\cal D}_{\alpha i} W_j^{\beta \, r}=  \delta_\alpha{}^\beta
\left(\Upsilon^r_{ij}- {1\over 2} \epsilon_{ij}   \Phi^I g_I^r\right) -{1\over 2} \epsilon_{ij} {\cal
F}^{ab\, r} (\gamma_{ab})_\alpha{}^\beta
 \; , \qquad
\end{eqnarray} as well as  the relations
 \begin{eqnarray}\label{DfcF=}
{\cal D}_{\alpha i} {\cal F}^r_{ab}= -2i\gamma_{[a|\alpha\beta}{\cal D}_{|b]} W_i^{\beta\, r} + i
\gamma_{ab\alpha}{}^{\beta}\Psi^I_{\beta i}g_I^r\; , \qquad \\ \label{DbcF=H} 3{\cal D}_{[c}{\cal
F}^r_{ab]}= {\cal H}^I_{abc}g_I^r\; . \qquad \end{eqnarray} Eq. (\ref{DfW=}) is equivalent to the
following set of equations
  \begin{eqnarray}\label{DW=Phi}
{\cal D}_{\alpha i} W^{\alpha i \, r}&=& 4 \Phi^I g_I^r \; , \qquad \\ \label{DW=Y} {\cal D}_{\alpha
(i} W_{j)}^{\beta \, r}&=& \delta_\alpha{}^\beta \Upsilon^r_{ij} \; , \qquad \\ \label{cF=gDW} {\cal
F}^r_{ab}&=& -{1\over 8}(\gamma_{ab})_\alpha{}^\beta {\cal D}_{\beta i} W^{\alpha i \, r}
 \;  \qquad
\end{eqnarray} and actually these are obtained from the Bianchi identities. Notice that Eqs.
(\ref{DW=Phi}) and (\ref{cF=gDW}) relate the superfields already present in (\ref{cF=ssp}) and
(\ref{cH3=ssp}) with higher components of the fermionic superfield $W^{\alpha i \, r}$, while
$\Upsilon^r_{ij}$  appears in (\ref{DW=Y}) as a notation for an irreducible component of ${\cal
D}_{\alpha i} W_j^{\beta \, r}$ which remains indefinite when studying the Bianchi identities.

Eq. (\ref{DbcF=H}) indicates that ${\cal F}_{ab}^r$ superfield is a (generalized) field strength of a
vector (super)field potential  and Eq. (\ref{DfcF=}) shows that all the higher components of ${\cal F}_{ab}^r$
are present in fermionic superfields entering (\ref{cF=ssp}) and (\ref{cH3=ssp}). Furthermore, using
(\ref{DfcF=})  in calculating the right hand sides ({\it r.h.s.}) of
  \begin{eqnarray}\label{DDW=gDW=}
\{ {\cal D}_{\alpha i}, {\cal D}_{\beta j}\} W_k^{\gamma \, r} &=&
2i\varepsilon_{ij}\gamma^a_{\alpha\beta}{\cal D}_a W_k^{\gamma \, r} =   \delta_\beta{}^\gamma \left(
{\cal D}_{\alpha i}\Upsilon^r_{jk} - {1\over 2} \epsilon_{jk}  {\cal D}_{\alpha i}  \Phi^I
g_I^r\right)+ ((\alpha i)\leftrightarrow (\beta j)) - \qquad \nonumber \\ &&  -{1\over 2} \epsilon_{jk}
{\cal D}_{\alpha i}{\cal F}^{ab\, r} (\gamma_{ab})_\beta{}^\gamma  + ((\alpha i)\leftrightarrow (\beta
j)) \;  \qquad \end{eqnarray} we find, after some algebraic manipulations,
   \begin{eqnarray}\label{DiracW=DY}
i(\gamma^a{\cal D}_a W_i^{r})_{\alpha} &=& {1\over 3}  {\cal D}_{\alpha}^j\Upsilon^r_{ij} +   {\cal
D}_{\alpha i}  \Phi^I g_I^r \; , \qquad \\ \label{sDY=0}
 {\cal D}_{\alpha (i}\Upsilon^r_{jk)}&=&0
\; , \qquad \\ \label{gDf=gPsi} {\cal D}_{\alpha i}  \Phi^I g_I^r &=&  2i\Psi_{\alpha i}^I g_I^r \; .
\qquad \end{eqnarray} This set of equations indicates that Eq. (\ref{cF=ssp}) itself describes the
off--shell constraints of the 6d (1,0) SYM model. Indeed, Eq. (\ref{DiracW=DY}) implies that the {\it
l.h.s.} of the Dirac equation for gaugino appears as a second component of the auxiliary superfield
$\Upsilon^r_{ij}$, the leading component of which is the auxiliary field of the SYM supermultiplet.

Taking a look from the other side, Eqs.  (\ref{DiracW=DY}) and (\ref{sDY=0}) can be collected in the following expression for the fermionic covariant derivative of the auxiliary superfield $\Upsilon^r_{ij}$
 \begin{eqnarray}\label{DfY=}
  {\cal D}_{\alpha i}\Upsilon^r_{jk}=2i \epsilon_{i(j} \left( \gamma^a{\cal D}_a W_{k)}^{r}
  - 2{\Psi}^I_{k)}   g_I^r \right)_{\alpha}  \; ,
\qquad \end{eqnarray}
which shows that higher components of the auxiliary superfield are expressed through the leading components of already introduced  basic superfields, {\it i.e.} through the fields of (1,0) superconformal theory of \cite{Samtleben:2011fj}.

Passing to  Binachi indentities (\ref{DcHI=}), from its lowest dimensional (dim 5/2) nontrivial
component   we find $0= \gamma^{a}_{\beta\gamma}\epsilon_{jk}(\eta_{ab}{\cal D}_{\alpha i}\Phi^I + 2i
\gamma_{ab\; \alpha }{}^\beta \Psi_{\beta i}^I)+ cyclic\; permutation \; of \; (\alpha i, \beta j,
\gamma k)$. Its general solution is
  \begin{eqnarray}\label{DfPhi=Psi}
  {\cal D}_{\alpha i}\Phi^I=2i \Psi_{\alpha i}^I
\; .
 \qquad
\end{eqnarray}

The dim 3 component of the Bianchi identity (\ref{DcHI=}) gives the following set of equations for the
fermionic derivative of the fermionic superfield $\Psi_{\alpha i}^I$:
\begin{eqnarray}
\label{ggDPsi=DPhi} \gamma_{ab\;(\alpha}{}^{\gamma} {\cal D}_{\beta ) (i}\Psi_{j)\gamma }^{I} &=& 2
id^I_{st} (W^s_i\gamma_{[a|})_{(\alpha|}(\gamma_{|b]}W^t_j)_{|\beta)}   \; , \quad \\
\label{gDPsi=DPhi}
 \tilde{\gamma}_{a}^{\beta\alpha} {\cal D}_{\alpha i}\Psi_{\beta }^{iI}&=& - 4{\cal D}_{a}\Phi^I \; ,
 \qquad  \\ \label{cHI=DPsi}
 {1\over 8} \tilde{\gamma}_{abc}^{\alpha\beta} {\cal D}_{\beta i}\Psi_{\alpha}^{iI}&=&  {\cal
 H}_{abc}^I +{i\over 4} d^I_{st}
W^{si}\gamma_{abc}W^t_i  \; , \qquad
 \qquad
\end{eqnarray}

Actually Eq. (\ref{cHI=DPsi}) gives more than that.  According to (\ref{6Dg3=sd}) the {\it r.h.s.} of
this equation  is anti-self dual while the second term in the {\it r.h.s.}  is  self-dual
 so that (\ref{6Dg3=sd}) can be decomposed onto
\begin{eqnarray} \label{cHI-=}
 {1\over 8}\tilde{\gamma}_{abc}^{\alpha\beta} {\cal D}_{\beta i}\Psi_{\alpha}^{iI}
= {\cal H}_{abc}^{- I}:= {1\over 2} ( {\cal H}_{abc}^{I}-  *{\cal H}_{abc}^{I}) \;  \qquad
 \qquad
\end{eqnarray} and \begin{eqnarray} \label{cHI+=0+}
 {\cal H}_{abc}^{+I}&:=& {1\over 2} ( {\cal H}_{abc}^{I}+  *{\cal H}_{abc}^{I})= - {i\over 4} d^I_{st}
W^{si}\gamma_{abc}W^t_i \; . \qquad
 \qquad
\end{eqnarray} Eq. (\ref{cHI+=0+}) is a generalization of the anti-self-duality conditions which
includes fermionic contributions. Thus studying the Binachi identities we have obtained a dynamical
equations from the superspace constraints (\ref{cH3=ssp}), (\ref{cH4=ssp}). Hence, in distinction to
(\ref{cF=ssp}) these latter are {\it on-shell} constraints.

Contracting Eq. (\ref{ggDPsi=DPhi}) with $\tilde{\gamma}^{c_1c_2c_3\;\alpha\beta} $, after some algebra
we find from its irreducible parts
\begin{eqnarray}\label{g3sDP} \tilde{\gamma}^{c\;\alpha\beta}{\cal
D}_{\alpha (i}\Psi_{j)\beta }^{I}=  2i W^s_{i}\gamma^cW^t_j\, d^I_{st} \; , \qquad
\tilde{\gamma}^{c_1c_2c_3\;\alpha\beta}{\cal D}_{\alpha (i}\Psi_{j)\beta }^{I}=0 \; . \qquad
\end{eqnarray} These equations imply \begin{eqnarray} \label{sDPsi=WgW} && {\cal D}_{\alpha
(i}\Psi_{j)\beta }^{I}=- {i\over 2} {\gamma}^{a}_{\alpha\beta } \, d^I_{st}\, W^{s}_i\gamma_{a}W^t_j \;
, \qquad  \end{eqnarray} Furthermore, Eqs. (\ref{gDPsi=DPhi}) and  (\ref{cHI-=}) can be written in the
form  of ${\cal D}_{[\alpha | i}\Psi_{|\beta ]}^{iI}= -{\gamma}^{a}_{\alpha\beta}{\cal D}_{a}\Phi^I$
and ${\cal D}_{(\alpha | i}\Psi_{|\beta )}^{iI}=- {1\over 6} {\gamma}^{abc}_{\alpha\beta } {\cal
H}_{abc}^{-I}$ so that
\begin{eqnarray} \label{DfiPhii=} {\cal D}_{\alpha  i}\Psi_{\beta }^{iI}=
-{\gamma}^{a}_{\alpha\beta}{\cal D}_{a}\Phi^I - {1\over 6} {\gamma}^{abc}_{\alpha\beta } {\cal
H}_{abc}^{-I} \end{eqnarray} Eqs. (\ref{DfiPhii=}) and (\ref{sDPsi=WgW}) together imply
\begin{eqnarray} \label{DfiPsij=} {\cal D}_{\alpha  i}\Psi_{\beta j}^{I}&=& {1\over 12}\epsilon_{ij}
{\gamma}^{abc}_{\alpha\beta }{\cal H}_{abc}^{-I}  + {1\over 2}\epsilon_{ij}
{\gamma}^{a}_{\alpha\beta}{\cal D}_{a}\Phi^I -  {i\over 2} {\gamma}^{a}_{\alpha\beta }
W^{s}_i\gamma_{a}W^t_j d^I_{st} \; , \qquad  \end{eqnarray} where in the first term the superscript
$^-$ can be actually omitted as far as  only the anti-self dual part of the 3-rank field strength
contributes to ${\gamma}^{abc}_{\alpha\beta } {\cal H}_{abc}^{I}$.

The dim 7/2 component of (\ref{DcHI=}) determines the fermionic derivative of the 3-rank antisymmetric
tensor superfield, \begin{eqnarray}   \label{DfcH3=} {\cal D}_{\alpha i}{\cal H}_{abc}^{I}= 3
i{\gamma}_{[ab|}{}_{\alpha}{}^{\beta }{\cal D}_{|c]}\Psi^I_{ \beta i} + 6 i{\cal F}^s_{[ab}
(\gamma_{c]}W_i^t)_{\alpha} d^I_{st}  - i{\gamma}_{abc}{}_{\alpha\beta} W_i^{\beta s} \, \Phi^J\,
d_{Jsr}g^{Ir} \; . \end{eqnarray} Finally, the dim 4 component of (\ref{DcHI=}) reads \begin{eqnarray}
\label{DbcH3=} {\cal D}_{[a}{\cal H}^I_{bcd]}={3\over 2 }d^I_{st}{\cal F}^s_{[ab}{\cal F}^t_{cd]}+
{1\over 4}g^{rI}{\cal H}_{abcd\, r}\; . \end{eqnarray} and implies that ${\cal H}_{abc}^{I}$ is a
generalized field strength.

Now let us observe that (\ref{DfiPsij=}) implies
  \begin{eqnarray}\label{DDPsi=gDPsi=}
\{ {\cal D}_{\alpha i}, {\cal D}_{\beta j}\} \Psi_{\gamma \, k}^I &=&
2i\varepsilon_{ij}\gamma^a_{\alpha\beta}{\cal D}_a \Psi_{\gamma \, k}^I= \qquad  \nonumber \\  &=&
{1\over 12} \epsilon_{jk} (\gamma^{abc})_{\beta\gamma}  {\cal D}_{\alpha i} {\cal H}^{-I}_{abc} - i
(\gamma^{a})_{\beta\gamma } {\cal D}_{\alpha i}W_{(j}^r \gamma^a W_{k)}^s d^I_{rs} + (\alpha i
\leftrightarrow \beta j) + \nonumber \\  && + i\epsilon_{jk} (\gamma^{a})_{\beta\gamma}  {\cal D}_{a}
\Psi_{\alpha i}^I  + i \epsilon_{ij} \epsilon_{\alpha\beta \gamma\delta} W^{\delta r}_k
\Phi^JX_{rJ}{}^I + (\alpha i \leftrightarrow \beta j) \; .  \qquad \end{eqnarray} To simplify the terms
in the last line we have used the commutation relation
 \begin{eqnarray}\label{DfDbPsi-=}
 {} [{\cal D}_{\alpha i}, {\cal D}_{a}] \Phi^I= i(\gamma_aW_i^r)_{\alpha}\Phi^JX_{rJ}{}^I :=
  2i(\gamma_aW_i^r)_{\alpha}\Phi^J (g_J^sd^I_{rs}- g^{sI} d_{Jrs})
\; ,    \qquad \end{eqnarray} Eq. (\ref{DfPhi=Psi}) as well as identities (\ref{6Dgammas}) and
$\epsilon_{ij}\epsilon^{kl}=-2 \delta_{i}^{[k} \delta_{j}^{l]}$.

Substituting the expressions (\ref{DfcH3=}) and (\ref{DfW=}) for the fermionic derivatives of
superfields in the {\it r.h.s.} of (\ref{DDPsi=gDPsi=}), after some algebra we obtain an equation one
of the irreducible parts of which ($\propto \epsilon_{ij}\epsilon_{\alpha\beta\gamma\delta} $) provides
us with superfield generalization of Dirac equations for the fermions of tensorial multiplet,
\begin{eqnarray}\label{gDPsi=}
 (\tilde{\gamma}^a{\cal D}_a \Psi_{k}^I)^{\delta}= {1\over 2} (\tilde{\gamma}^{ab}W_i^r)^{\delta}
 d^I_{rs} {\cal F}^s_{ab}  + \Upsilon^s_{ij} W^{\delta j r} d^I_{rs} +{1\over 2} W_i^{\delta r}\Phi^J
 (-g_J^sd^I_{sr}+4 d_{Jsr}g^{Is})
\; .     \qquad \end{eqnarray}
The other irreducible parts of the above mentioned equation (symmetric
in $(\alpha\beta)$ and $(ij)$) are satisfied identically due to  $\gamma^c
\tilde{\gamma}{}^{def} \gamma_{ab} \gamma_c =-4 \gamma_{[a} \tilde{\gamma}{}^{def} \gamma_{b]}$, which
one can easily prove using the gamma matrix algebra.

Now let us turn to the Bianchi identities (\ref{DcH4=}). To this end we need the constraints for the
5-form  ${\cal H}_{5A}$ which we assume to be
\begin{eqnarray}   \label{cH5=ssp} {\cal H}_{5A}&=&
{1\over 4!} E^d\wedge E^c\wedge E^b\wedge E^a \wedge E^{\alpha i} {\cal H}_{\alpha i\; abcd\; A} +
{1\over 5!} E^e\wedge E^d\wedge E^c\wedge E^b\wedge E^a {\cal H}_{ abcde\; A}  \qquad \end{eqnarray}
Then dim 7/2 and lower  components of (\ref{DcH4=}) are satisfied identically, \footnote{Actually, the
dim 7/2 component  can be used to fix $l=-1/6$ if we start with the constraint (\ref{cH3=ssp}) with
coefficient $il$ for the first term, as it was discussed in the footnote 1.} while its dim 4 component,
after some algebra,  can be presented in the form
\begin{eqnarray}\label{DcH4=bbbff} &0&=
i\epsilon_{ij}\gamma^d_{\alpha\beta} \left({\cal H}_{abcd\; r}-{1\over 2}\epsilon_{abcdef}{\cal
F}^{ef\; s} \Phi^I d_{Irs} -{i\over 2}\epsilon_{abcdef} (W^{ks}\gamma^{ef}\Psi^I_k)d_{Irs}\right)
-\qquad \nonumber \\ && - i \gamma_{abc \alpha\beta}(\Upsilon_{ij}^s\Phi^Id_{Irs} -2i W^s_{(i}
\Psi^I_{j)}d_{Irs})\; .  \nonumber \\ \end{eqnarray} Clearly, the first and the second line of Eq.
(\ref{DcH4=bbbff}) belong to different irreducible representations of $SO(1,5)$ and thus vanish
separately. Hence  we have found the duality equation deformed by fermionic contribution,
\begin{eqnarray}\label{cH4=Fpd} {\cal H}_{abcd\; r}-{1\over 2}\epsilon_{abcdef}{\cal F}^{ef\; s}\Phi^I
d_{Irs}= {i\over 2}\epsilon_{abcdef} (W^{ks}\gamma^{ef}\Psi^I_k)d_{Irs}  \; \end{eqnarray} and the superfield
generalization of the equations for auxiliary scalar field of the SYM multiplet, also involving the
fermionic superfields,
\begin{eqnarray}\label{Yphd=ff} {\cal E}_{ij\; r}:= (\Upsilon_{ij}^s\Phi^I - 2i
W^s_{(i} \Psi^I_{j)})d_{Irs}=0 \; .
\end{eqnarray}
The next-to leading component of this auxiliary superfield equation gives the fermionic equation of motion. Indeed, after some algebra one finds that  the symmetric in $(ijk)$ part of the equation ${\cal D}_{\alpha k} {\cal E}_{i j\; r}=0$ is satisfied identically due to the first equation in (\ref{conaction}), while the remaining irreducible part, $\epsilon^{jk}{\cal D}_{\alpha k} {\cal E}_{i j\; r}=0$, reads
\begin{eqnarray}\label{PdDgW=}
\Phi^I
d_{Irs}(\gamma^a{\cal D}_a W_i^{s})_{\alpha} &=& - {1\over 12}(\gamma^{abc} W_i^{s})_{\alpha} {\cal H}_{abc} ^{-I}d_{Irs}-{1\over 2} (\gamma^aW_i^{s})_{\alpha}  {\cal D}_a\Phi^I
d_{Irs} - \qquad \nonumber \\ && - {1\over 2} (\gamma^{ab}\Psi_i^I)_{\alpha}{\cal F}_{ab}^{s}
d_{Irs} +   {1\over 2}\Psi^I_{\alpha\, i}  \Phi^J
\left(4 g_I^s d_{Jrs}- g_J^s d_{Irs} \right) - \qquad \nonumber \\ &&  -
 \Upsilon^s_{ij}\Psi^I_{\alpha\, i}   d_{Irs} + {2i\over 3}\epsilon_{\alpha\beta\gamma\delta} W^{\beta j\, s }W_j^{\gamma u}W_{i}^{ \delta v} d^I_{rs}d_{Iuv}
  \; . \qquad \end{eqnarray}
The same equation can be obtained from (\ref{DiracW=DY}) by multiplying it by $\Phi^I
d_{Irs}$  and using Eqs. (\ref{Yphd=ff}), (\ref{DfW=}) and (\ref{DfiPsij=}) as well as the first equation in (\ref{conaction}).

Dim 9/2 component of (\ref{DcH4=}) gives the expression for the fermionic derivative of the 4th rank
antisymmetric tensor superfield, \begin{eqnarray}   \label{DfcH4=} {\cal D}_{\alpha i} {\cal H}_{abcd\,
r}&=&  8i {\cal H}^I_{[abc} (\gamma_{d]}W^s_i)_\alpha d_{Irs} - 12i {\cal F}^s_{[ab}
(\gamma_{cd]}\Psi^I_i)_\alpha d_{Irs} + 4i  {\cal D}_{[a}({\Phi}^I(\gamma_{bcd]}W^{s}_i)_{\alpha} )
d_{Irs} + \quad \nonumber \\ && + {\cal H}_{\alpha i\, abcd\, A}k^A_{ r} \; ,  \qquad \end{eqnarray}
and the dim 5 component states that this tensorial superfield is a generalized field strength
\begin{eqnarray}   \label{DbcH4=} {\cal D}_{[a} {\cal H}_{bcde]\, r}= - 4{\cal F}^s_{[ab} {\cal
H}^I_{cde]} d_{Irs}+{1\over 5} {\cal H}_{abcde\, A}k_r^A\; .  \qquad \end{eqnarray}

One can state that Eq. (\ref{DfcH4=}) completely determines the  fermionic  derivative of the 4th rank
antisymmetric tensor superfield ${\cal H}_{abcd\, r}$ only if we express ${\cal H}_{\alpha i\, abcd\,
A}$ in terms of already known superfields. To do this,   it is convenient  to study the Bianchi
identities for the 5-form field strength. According to  \cite{Samtleben:2011fj} they read
\begin{eqnarray}\label{DcH5=} I_{6A}:= D{\cal H}_{5A} + c_{A\,IJ}  {\cal H}_3^I \wedge  {\cal H}_3^J +
c_{A\,s}^r {\cal F}^s \wedge{\cal H}_{4r}+ \ldots =0 \; , \qquad \end{eqnarray} where $\ldots $ denote
the possible term which vanish when contracted with $k_r^A$. The consistency condition for the Bianchi
identity (\ref{DcH5=}),
'identity for indentity'\footnote{Actually the 7-form 'identity for identity' $J_{7\, A}=0$ reduces to $DI_{6A}=0$ when the lower form Bianchi identities are  satisfied. See Appendix C for its complete  form.}
\begin{eqnarray}\label{IdfId5=}
DI_{6A}=0 \;  \qquad
\Leftrightarrow \quad
DD {\cal H}_{5A}= - {\cal H}_{5B}\wedge {\cal F}^r\,
X_{rA}{}^B
 \qquad \end{eqnarray}
 implies that
  \begin{eqnarray}\label{XrAB=}X_{rA}{}^B= c_A{}^s{}_r k^B_s  \;,\qquad
c_A{}^t{}_{(s|}  d_{I\, |r)t}  = c_{A\, IJ} d^J_{rs}\;,\qquad  c_A{}^r{}_s g_{Ir}  = -2 c_{A\, IJ}
g^{J}_s \; . \qquad \end{eqnarray} With our constraints the first nontrivial component equation in
(\ref{DcH5=}) has dimension 9/2. It can be solved by \begin{eqnarray}\label{9/2DcH5=}
  {\cal H}_{\alpha i\, abcd\; A}= i(\gamma_{abcd}\Psi^J_i)_{\alpha}\Phi^I c_{AIJ}
= {i\over 2} \epsilon_{abcdef}(\gamma^{ef}\Psi^J_i)_{\alpha}\Phi^I c_{AIJ} \; . \qquad \end{eqnarray}
Then the next, dim 4 component  of (\ref{DcH5=}) produces a supersymmetric generalization of the
duality relation between 4 form potential and scalar,
\begin{eqnarray}\label{cH5=}
  {\cal H}_{abcde\; A}= {1\over 2} \epsilon_{abcdef}\left(c_{A\, IJ}\left(\Phi^{[I}{\cal
  D}^f\Phi^{J]}-2i \Psi^{iI}\tilde{\gamma}^f\Psi^{J}_i\right)-ic_A{}^r{}_{[s}d_{t]r\, I}\Phi^I
  W^{is}\gamma^fW_i^t\right)
\; . \qquad \end{eqnarray}

As we have already mentioned, (\ref{XrAB=}) should be obeyed modulo terms which vanish when contracted
with $k_r^A$. So, if we consider constraints (\ref{cH5=ssp}) contracted with $k_r^A$, their consequence
will contain the duality equation \begin{eqnarray}\label{k4cH5=}
  k_r^A{\cal H}_{abcde\; A}= {1\over 2} \epsilon_{abcdef} \left( \left(\Phi^{[I}{\cal
  D}^f\Phi^{J]}-2i\Psi^{iI}\tilde{\gamma}^f\Psi^{J}_i\right) k_r^A c_{A\, IJ}
- i k_r^A  c_A{}^u{}_{[s}d_{t]u\, I}\Phi^I W^{is}\gamma^fW_i^t \right).   \nonumber \\ {} \end{eqnarray}

Notice that another  $SO(1,6)\times SU(2)$ irreducible part of the  dim 4 component  of (\ref{DcH5=})
(the one $\propto \epsilon_{ij}\gamma_{\alpha\beta [a} (\ldots)_{bcd]}$) results in
\begin{eqnarray}\label{Wg3Wkcd=0} \left({\cal H}_{bcd}^{+J} +{i\over 4} W^{ks}\gamma_{bcd}W^t_k\,
d^J_{st}\right) c_{AIJ}\Phi^I=0 \; ,  \; \end{eqnarray} which is obeyed identically due to the
supersymmetrized anti-self duality equation (\ref{cHI+=0+}).

To obtain the first order bosonic equations (duality and anti-self-duality conditions) derived from the closure of the supersymmetry algebra in
\cite{Samtleben:2011fj} (and, in their purely bosonic limit, from the action in \cite{BSS2013}), we
have to identify the leading components of the antisymmetric tensor superfield with field strengths,
$\Upsilon^r_{ij}\vert_{\theta=0}=Y_{ij}^r$ and $\Phi^I\vert_{\theta=0}=2\phi^I$, where the only
nontrivial coefficient appears.

The second order bosonic equations can be obtained from the (self)duality relations
 and the purely bosonic higher dimensional components of the  Bianchi identities. On the other hand, they can be obtained from the next-to leading components of the superfield fermionic equations. As far as the fermionic equations, in their turn, can be obtained by acting on the duality equations by fermionic covariant derivatives, the coincidence of the second order bosonic equations obtained in this two ways provides an additional check of the consistency of our constraints  or, in other words, of the equivalence of the superspace constraints and the spacetime component equations of (1,0) superconformal theory of \cite{Samtleben:2011fj}.
Actually it is a superspace counterpart of checking the closure of supersymmetry algebra on the spacetime component equations, which was done in \cite{Samtleben:2011fj}.
Below we perform a  bit simplified version of this consistency check, following mainly the bosonic superfield  contributions to the second order bosonic equations.

\subsection{Second order bosonic equations and check of consistency of our superspace description of the (1,0) superconformal theory}

To begin the final check of consistency of our superspace description, let us discuss how the fermionic equations can be obtained from the duality and anti-self-duality conditions.

\subsubsection{Fermionic equations of motion  from duality and anti-self-duality conditions}

The tensor multiplet fermionic equation provided by the leading component of the superfield equation  (\ref{gDPsi=}),  which we denote by  ${\cal E}_{{ i}\,}^{\delta  I}=0$, appears also as the next-to leading component of the generalized anti--self--duality condition (\ref{cHI+=0+}), which we denote by ${\cal I}_{ abc}^{+I}=0$.
Indeed, acting on this latter superfield equation  by ${\cal D}_{\alpha i}$, after some algebra one finds that  the only nontrivial irreducible part of   ${\cal D}_{\alpha i}{\cal I}_{ abc}^{+I}=0$ is $\tilde{\gamma}^{abc\, \delta \alpha}{\cal D}_{\alpha i}{\cal I}_{ abc}^{+I}=0$, which coincides with Eq. (\ref{gDPsi=}). The other irreducible parts of ${\cal D}_{\alpha i}{\cal I}_{ abc}^{+I}=0$ are satisfied identically:  after the use of Eq. (\ref{DfW=}) and of the Bianchi identity  (\ref{DfcH3=}), which we denote by $I_{\alpha i abc}^I=0$, one finds that
${\cal D}_{\alpha i}{\cal I}_{ abc}^{+I} = i {\gamma}_{abc \alpha\beta} {\cal E}_{{i}\,}^{\beta I}$.

Similarly, the vector multiplet  fermionic superfield equation (\ref{PdDgW=}),  ${\cal E}_{\beta i\; r}=0$, can be obtained from the duality equation (\ref{cH4=Fpd}), ${\cal I}_{c_1...c_4\, r}=0$. This fact is expressed by the following form of the 5-form Bianchi identity with all but one bosonic indices (\ref{DfcH4=}):
 \begin{eqnarray}
 \label{Ifbbbbr=}I_{\alpha i\; c_1...c_4\, r}= {\cal D}_{\alpha i} {\cal I}_{c_1...c_4\, r}
+{1\over 2}\epsilon_{abc_1...c_4}I_{\alpha i}{}^{ab \; s}\Phi^Id_{Isr}-{i\over 2}\epsilon_{abc_1...c_4}\gamma^{ab}{}_\alpha{}^\beta {\cal E}_{\beta i\; r} =0 \; .  \qquad
\end{eqnarray}
Here $I_{\alpha i\, ab}{}^{s}=0$ is the dim 5/2 component of the gauge field Bianchi identity, Eq. (\ref{DfcF=}).

One can also check that no new equations appear when acting by fermionic covariant derivative on the 5-form--scalar duality  $k_r^A {\cal I}_{c_1...c_5\, A}=0$ (\ref{k4cH5=}) and using the
dim 11/2 Bianchi identity
\begin{eqnarray}
 \label{kI=11-2}
k_r^A {I}_{\alpha i c_1...c_5\, A}:=k_r^A {D}_{\alpha i}{\cal H}_{c_1...c_5\, A}+\ldots = 0\; .  \qquad
\end{eqnarray}

To resume, we have shown that the fermionic superfield equations (\ref{PdDgW=}) and (\ref{gDPsi=}) can be obtained (also) by acting by fermionic covariant derivatives on the duality and anti-self-duality conditions. No additional restrictions on the physical fields, beyond the equations of (1,0) superconformal theories of \cite{Samtleben:2011fj} are produced at this stage.

\subsubsection{Second order equation for 3-form (super)field strength }

The second order equation for the 3-form field strength ${\cal H}^I_{abc}$ can be found calculating the {\it r.h.s.} of the identity ${\cal D}_c{\cal H}^{abc I}= -{1\over 3!}\epsilon^{abcdef}{\cal D}_{c}{\cal H}_{def}^{I}+ 2 {\cal D}_c{\cal H}^{abc\, +I}$ with the use of the generalized anti-self-duality equation
(\ref{cHI+=0+}) and the tensorial Bianchi identities (\ref{DbcH3=}). In such a way one arrives at
\begin{eqnarray} \label{DbcHI=Eqs}
{\cal E}^{abI}:= {\cal D}_c{\cal H}^{abc I} &+& {1\over 4}\epsilon^{abcdef}{\cal F}_{cd}^r {\cal F}_{ef}^s d^I_{rs} + {1\over 4!}\epsilon^{abcdef}{\cal H}_{cdef\, r}g^{rI}  + i{\cal D}_cW^{si}\gamma^{abc}W^t_i   d^I_{st} =0  \; . \;
 \qquad
\end{eqnarray}
Furthermore, using (\ref{cH4=Fpd}) we can present this equation in the form
  \begin{eqnarray} \label{DbcHI=Eqs=3}
{\cal E}^{abI}:= {\cal D}_c{\cal H}^{abc I} &+& {1\over 4}\epsilon^{abcdef}{\cal F}_{cd}^r {\cal F}_{ef}^s d^I_{rs} - {\cal F}^{ab\, r}\Phi^J d_{Jrs}g^{rI} + \qquad \nonumber \\ && +i{\cal D}_cW^{si}\gamma^{abc}W^t_i   d^I_{st} + i W^{it}\gamma^{ab}\Psi^J_ig_J^sd^I_{st}=0  \; .
 \qquad
\end{eqnarray}

The second order bosonic equations can be also obtained form the next-to leading components of the superfield fermionic equations. The coincidence of the results of these two calculations provides an additional check of the equivalence of our superspace constraints and of the spacetime component formalism of \cite{Samtleben:2011fj}:  at this stage some extra restrictions on our fields, beyond the spacetime component equations for the fields of (1,0) superconformal theory, could appear if the constraints were too strong.

Acting by fermionic covariant derivative on the fermionic equations (\ref{gDPsi=}), which we denote by ${\cal E}_{{ i}\,}^{\delta  I}=0$, one finds that ${\cal D}_{\alpha i}{\cal E}_{{ j}\,}^{\beta I}=
{i\over 4} \tilde{{\cal E}}{}^{ab I} \gamma_{ab}{}_{\alpha}{}^\beta -
{i\over 2} {\cal E}^{I} \delta_{\alpha}{}^\beta - 2i \delta_{\alpha}{}^\beta {\cal E}_{ij\, r}g^{r\, I}$, where $ {\cal E}_{ij\, r}=0$ is the auxiliary superfield equation (\ref{Yphd=ff}), ${\cal E}^{I} =0$ is the scalar superfield equation
\begin{eqnarray} \label{DDPI=Eqs}
{\cal E}^{I}:= \Box{}\Phi^I  - {\cal F}_{ab}^r {\cal F}^{ab\, s} d^I_{rs} &-& {3\over 2}\Phi^Jg_J^r\Phi^Kg_K^s  d^I_{rs}+ {\Upsilon}_{ij}^r {\Upsilon}^{ij\, s} d^I_{rs} + \qquad \nonumber \\ && + i{\cal D}_aW^{si}\gamma^{a}W^t_i   d^I_{st} + i W^{is}\Psi_i^J (4g_J^t d^{I}_{st}  - g^{It}d_{Jst} ) =0
 \; ,
 \qquad
\end{eqnarray}
and $\tilde{{\cal E}}{}^{ab I}=0$ reads
\begin{eqnarray} \label{DbcHI=Eqs=2}
\tilde{{\cal E}}{}^{ab I}:= {\cal D}_c{\cal H}^{abc I} + {1\over 4}\epsilon^{abcdef}{\cal F}_{cd}^r {\cal F}_{ef}^s d^I_{rs}-
  {1\over 4!}\epsilon^{abcdef}{\cal H}_{cdef\, r}g^{rI}
  - 2 {\cal F}^{ab\, r}\Phi^J d_{Jrs}g^{rI} + \qquad  \nonumber  \\ + fermions =0  . \quad
\end{eqnarray}
For simplicity, from now on we will mainly follow the bosonic superfield contributions to the second order bosonic (super)field equations.

On the first look, the tensorial equation  (\ref{DbcHI=Eqs=2}), $\tilde{{\cal E}}{}^{ab I}=0$,
differs from (\ref{DbcHI=Eqs}), ${{\cal E}}{}^{ab I}=0$,  as far as the fourth term in the  former is absent in the latter. However, a more close look permits to notice also the difference in the sign in front of  the third terms, $\pm {1\over 4!}\epsilon^{abcdef}{\cal H}_{cdef\, r}g^{rI} $, and to appreciate that actually these equations coincide modulo the duality equation (\ref{cH4=Fpd}), ${\cal I}_{abcd\, r}=0$. Resuming, \begin{eqnarray} \label{cE=cE+cI}\tilde{{\cal E}}{}^{ab I}= {{\cal E}}{}^{ab I}- {1\over 12}\epsilon^{abcdef}{\cal I}_{cdef\, r}g^{rI} . \quad
\end{eqnarray}

This relation between the forms of the second order equations obtained from the superfield fermionic equation and directly from the selfduality condition has an interesting consequence.
As far as the superfield fermionic equation (\ref{gDPsi=}) itself can be obtained by acting by the fermionic covariant derivatives on (\ref{cHI+=0+}) (see sec. 3.4.1), we can state that also the duality conditions (\ref{cH4=Fpd}) projected on $g^{rI}$, ${\cal I}_{cdef\, r}g^{rI}=0$, can be obtained from the generalized anti-self-duality superfield equation (\ref{cHI+=0+}).

\subsubsection{Second order equation for the gauge (super)field strength }

Similarly, from the duality equation (\ref{cH4=Fpd}) we  find
  \begin{eqnarray} \label{DbFPd=H5+}
{\cal E}^{b}_r:=  d_{Irs} {\cal D}_a(\Phi^I{\cal F}^{ab\, s})+{1\over 3!}
 \epsilon^{bcdefg} {\cal F}_{cd}^{s} {\cal H}_{ efg}^{I} d_{Irs}-{1\over 5!}
 \epsilon^{bcdefg} {\cal H}_{ cdefg\; A}k_r^{A} + \nonumber \\  + i
 {\cal D}_a (W^{is}\gamma^{ab}\Psi_i^Jd_{Jrs}) =0 \; , \qquad \end{eqnarray}
 and then, using (\ref{cH5=}),
 \begin{eqnarray} \label{DbFPd=}
{\cal E}^{b}_r &=&  d_{Irs} {\cal D}_a(\Phi^I{\cal F}^{ab\, s})+{1\over 3!}
 \epsilon^{bcdefg} {\cal F}_{cd}^{s} {\cal H}_{ efg}^{I} d_{Irs} -
{1\over 2} k_r^A c_{A\, IJ} \Phi^{[I}{\cal
  D}^b\Phi^{J]}+\qquad \nonumber \\ && + i k_r^A c_{A\, IJ}  \Psi^{iI}\tilde{\gamma}^b\Psi^{J}_i + {i\over 2} k_r^A c_A{}^u{}_{[s }d_{t]u I}\Phi^I
  W^{is}\gamma^fW_i^t  + i
 {\cal D}_a (W^{is}\gamma^{ab}\Psi_i^Jd_{Jrs})  =0
 \; .  \qquad \end{eqnarray}

On the other hand, let us consider the  equation ${\cal D}_{\beta j} {\cal E}_{\alpha i}^r =0$  obtained by acting by  the  fermionic covariant derivatives on the  fermionic superfield equation of motion (\ref{DiracW=DY}),  ${\cal E}_{\alpha i}^r =0$. Using
(\ref{DfW=}), (\ref{DfiPsij=}) and (\ref{sDY=0})
we find that the SU(2) tensorial part of this equation, ${\cal D}_{\beta (j} {\cal E}_{ i) \alpha}^r =0$, is satisfied identically,  while the
$SU(2)$ singlet part, $\epsilon^{ij}{\cal D}_{\beta j} {\cal E}_{\alpha i}^r =0$, gives rise to the self-dual part of the bosonic Bianchi identity (\ref{DbcF=H}) (from ${\cal D}^i_{(\beta} {\cal E}_{\alpha ) i}^r =0$) and to
\begin{eqnarray}\label{DaFab=}
{\cal D}^a{\cal F}_{ab}^r+ {1\over 2}  {\cal D}_b\Phi^Ig_I^r + {i\over 2} W^{is}\gamma_{bcd}W^t_k\,
f^r_{ts} +{i\over 4!}
\tilde{\gamma}{}^{\alpha\beta}_b{\cal D}_{\alpha}^i{\cal D}_{\beta}^j \Upsilon^r_{ij}=0 \; .  \; \end{eqnarray}
Although formally this looks like (the superfield generalization of) the interacting gauge field equation of motion it contains the term with auxiliary superfield $\Upsilon^r_{ij}$. This reflects the off--shell nature of the SYM part of our constraints. In the interacting system the other constraints result in that  $ \Upsilon^r_{ij}$ must be a solution of  the algebraic equation (\ref{Yphd=ff}). However, to use this equation, and thus to make Eq. (\ref{DaFab=}) dynamical, we should multiply it by $\Phi^Id_{Irs}$. Then, using the Leibnitz rule to move the fermionic covariant derivatives in the last term of this equation, as well as Eqs.
(\ref{Yphd=ff}), (\ref{DfW=}), (\ref{DfiPsij=}), (\ref{DfcF=}), (\ref{DfPhi=Psi}) and (\ref{DiracW=DY}), after some algebraic manipulation we arrive at
\begin{eqnarray}\label{DaPhiFab=}
 0= d_{Irs} {\cal D}^a(\Phi^I{\cal F}_{ab}^r)+ {1\over 3!}
 \epsilon_{bcdefg} {\cal F}^{cd\, r} {\cal H}^{- \, efg\, I} d_{Irs}-  \Phi^I {\cal D}_b\Phi^J g_{[I}^r d_{J]rs}  + fermions  \; , \qquad \end{eqnarray}
Notice that the above mentioned transformations have resulted in  appearance of the second  term $\propto \epsilon {\cal F}^{r} {\cal H}^{- \,  I} d_{Irs}$, absent in (\ref{DaFab=}),  and in antisymmetrizing the  indices of the product of invariant tensors in the coefficient for the scalar current, $g_{J}^r d_{Irs}  \mapsto g_{[J}^r d_{I]rs}$. Now, the terms presented in  (\ref{DaPhiFab=}) coincide with the ones in  (\ref{DbFPd=}) because of the property $g_{[I}^r d_{J]rs}= {1\over 2}k_s^Ac_{AIJ}$  (see  (\ref{conaction})), and as far as, due to (\ref{cHI+=0+}), ${\cal H}^{efg\, I}= {\cal H}^{- \, efg\, I}+ fermions$.

\subsubsection{Scalar (super)field equation and 5-form duality condition}

The second order equation for scalar superfield (\ref{DDPI=Eqs}) has been obtained above form the fermionic superfield equation of motion (\ref{gDPsi=}). On the other hand, let us consider the duality equation (\ref{k4cH5=}), which we denote by ${\cal I}_{c_1...c_5\, A}k_r^A=0$. Taking the covariant divergence of its Hodge dual,
${1\over 5!}\epsilon^{bc_1...c_5}{\cal I}_{c_1...c_5\, A}k_r^A$, and using the pure bosonic part of the  Bianchi identities (\ref{DcH5=}),
\begin{eqnarray}\label{DcH5=6}
{1\over 5!}\epsilon^{bc_1...c_5}D_b{\cal H}_{c_1...c_5\, A}k_r^A=  {1\over 3! \cdot 3!}
 \epsilon^{bcdefg} {\cal H}_{bcd}^{I}{\cal H}_{ efg}^{J} k_r^A c_{A\,IJ} + {1\over 2! \cdot 4!} k_r^Ac_{A\,s}^r \epsilon^{abc_1...c_4} {\cal F}^s_{ab} {\cal H}_{c_1...c_4\,r} ,  \nonumber \\
  \end{eqnarray}
as well as Eqs. (\ref{cH4=Fpd}) and (\ref{cHI+=0+}), we arrive at
\begin{eqnarray}\label{eDcH5=6}
0=  {1\over 2} k_r^A c_{A\,IJ} \Phi^I\left( \Box{}\Phi^J - d^J_{st} {\cal F}^s_{ab}  {\cal F}^{ab\, t}\right) + fermions
\; . \qquad
  \end{eqnarray}
Notice that $\epsilon^{bcdefg} {\cal H}_{bcd}^{I}{\cal H}_{ efg}^{J}= \epsilon^{bcdefg} {\cal H}_{bcd}^{[I}{\cal H}_{ efg}^{J]} =fermions$. This is the case  due to Eq. (\ref{cHI+=0+}), which implies ${\cal H}_{bcd}^{I}= {\cal H}_{bcd}^{-I}+fermions$, and the identity
$\epsilon^{bcdefg} {\cal H}_{bcd}^{-[I}{\cal H}_{ efg}^{J]-}=- 6 {\cal H}_{bcd}^{-[I}{\cal H}^{J]-bcd}\equiv 0$.

Substituting (\ref{DDPI=Eqs}) and following, for simplicity, only the contributions of the bosonic superfields, one finds that (\ref{eDcH5=6}) is equivalent to
 $$ 0=  - {3\over 2}k_r^A c_{A\,IJ}\Phi^Jg_J^r\Phi^Kg_K^s  d^I_{rs}+ {1\over 2} k_r^A c_{A\,IJ} {\Upsilon}_{ij}^r {\Upsilon}^{ij\, s} d^I_{rs}+fermions\, . $$
 Using the properties of the invariant tensors in (\ref{conaction}) one can find that this is indeed the case; namely, the first term in this equation vanishes identically, while
the second is expressed through the fermionic bilinears   with the use of auxiliary superfield equation (\ref{Yphd=ff}).

Thus  we have obtained the second order bosonic equations by acting by bosonic derivatives on the  duality conditions, and also, following mainly the bosonic superfield contributions, by acting by fermionic derivative on the fermionic superfield equations. The bosonic equations obtained on these two ways are equivalent; no additional restrictions on physical fields appears. This procedure is a superfield counterpart of searching for closure of supersymmetry algebra on the spacetime component equations performed in \cite{Samtleben:2011fj}.

\subsection{Summary}

Thus we have performed the complete investigation of the Bianchi identities (\ref{DcF=}), (\ref{DcHI=}), (\ref{DcH4=}), (\ref{DcH5=}) with our superspace constraints (\ref{cF=ssp}), (\ref{cH3=ssp}), (\ref{cH4=ssp}), (\ref{cH5=ssp}), (\ref{9/2DcH5=}), have studied the consequences of this solution and found that our superspace constraints describe the  (1,0) superconformal theory of \cite{Samtleben:2011fj}.

All the  physical fields of this (1,0) superconformal theory appear  as leading components of the fermionic and bosonic main superfields, $W^{\alpha r}_i$, $\Psi_{\alpha i}^I$ and $\Phi^I$, ${\cal F}_{ab}^r$, ${\cal H}_{abc}^I$, ${\cal H}_{abcd\, r}$,
${\cal H}_{abcde\, A}$, which enter the differential form representation of our constraints (\ref{cF=ssp})--(\ref{cH5=ssp}), (\ref{9/2DcH5=})\footnote{Actually, the meaning of the term 'main superfield' in the superspace literature is usually more restrictive, but the wider treatment of it in this section cannot lead to any confusion.}. The auxiliary field of the SYM multiplet enters as leading component in the auxiliary superfield $\Upsilon^r_{ij}$  which appears as one of the irreducible parts of the fermionic derivative of the $W^{\alpha r}_i$.

The complete solution of the Bianchi identities, which we have described above, gives us the relations between main superfields, including the anti-self-duality and duality relations (\ref{cHI+=0+}), (\ref{cH4=Fpd}) and (\ref{k4cH5=}), algebraic auxiliary superfield equation (\ref{Yphd=ff}), as well as the  expressions for the covariant derivatives of the main superfields.

The next stage consisted in obtaining the consequence of our solution, which has been done by studying the results of the action of fermionic covariant derivatives on the above described relations between the main superfields and their fermionic derivatives. (Actually, we did this for all the relations but the superfield generalization  of pure bosonic tensorial Bianchi identities which are dependent as we discuss in Appendix C). We have  obtained the superfield generalization of the fermionic equations of motion (\ref{PdDgW=}) and (\ref{gDPsi=})  by acting on the duality and self duality relations by fermionic covariant derivatives. The superfield generalization of the second order bosonic equations can be obtained by acting by the bosonic derivative on the duality and anti-self-duality equations and by acting by the fermionic covariant derivatives on the fermionic superfield equations. We have shown  (sometimes for simplicity following the bosonic superfield contributions on one of two ways) that the purely bosonic equations obtained on these two ways coincide.
This final check can be considered as a superfield counterpart of the closure of supersymmetry algebra on the equations.

To resume, our superspace constraints on the field strengths of the tensorial hierarchy (\ref{cF:=})--(\ref{DcH4=}) restrict the field content of the superfields to the fields  of the (1,0) superconformal theory of  \cite{Samtleben:2011fj}, produce exactly the same equations of motion for the physical  fields  as were obtained from the closure of (1,0) supersymmetry algebra in \cite{Samtleben:2011fj}, do not produce other restrictions on the physical fields of the (1,0) superconformal theory  and, hence, are equivalent to the spacetime component equations of motion of
this found in \cite{Samtleben:2011fj}.

\section{Conclusions}

Thus we have shown that all the dynamical equations for the bosonic fields of the 6D (1,0) superconformal
theories of \cite{Samtleben:2011fj} can be obtained from the superspace constraints (\ref{cF=ssp}),
(\ref{cH3=ssp}), (\ref{cH4=ssp}) and (\ref{cH5=ssp}). Instructively, these dynamical equations were
obtained in the form of superfield duality and anti-self-duality conditions (\ref{cHI+=0+}), (\ref{cH4=Fpd}) and (\ref{k4cH5=}).

We have shown that the superfield generalization of the fermionic equations can be obtained from these first order duality equations by acting  by the fermionic covariant derivatives, while a suitable action by   the bosonic covariant derivative produces the (superfield generalization of the) second order bosonic field equations.  Following mainly the bosonic superfield contributions, we have also obtained the above mentioned second order bosonic equations by acting by fermionic derivatives on the fermionic superfield equations. This has been  an additional consistency check, designed to convince the reader that our superspace constraints do not impose any additional condition on  the fields of the (1,0) superconformal
theory, but only the spacetime field equations  of \cite{Samtleben:2011fj}, has confirmed  that our superfield formalism is equivalent to the spacetime component description of the 6D (1,0) superconformal theory developed in  \cite{Samtleben:2011fj}.

The superspace realization of the tensorial hierarchy of \cite{Samtleben:2011fj} developed in this
paper clarifies the structure and provides a new look on the (1,0) superconformal theory of
\cite{Samtleben:2011fj}. It can be useful in the search \cite{BSS2013+} for  the supersymmetric
generalization of the purely bosonic action of \cite{BSS2013} and  hopefully, can provide an insight in
the quest for a hypothetical (2,0) superconformal theory related to multiple M5--brane system.
Probably to this end it will be useful to understand better the possible relation of our constraints with the 6d twistor approach of \cite{Saemann+,Palmer:2013pka}.

\bigskip

{\bf Acknowledgments}. {The author is thankful to Dima Sorokin and Henning Samtleben for collaboration
on related projects \cite{BSS2013,BSS2013+} and encouraging comments.
This work was supported in part
by the research grant FPA2012-35043-C02-01 from the MEC of Spain, by the Basque
Government Research Group Grant ITT559-10 and by the UPV/EHU under the program UFI 11/55. The hospitality and support of  the Theoretical Department of CERN at very final stages of this work is greatly acknowledged.
}

\begin{appendix}

\section{Algebraic constraints on the constant tensors} \label{app:con}

The consistency conditions for the Bianchi identities of the  tensorial hierarchy
(\ref{DcF=})--(\ref{DcH4=}) require the  tensors  $f_{st}{}^r$, $d^I_{rs}$, $g^{Ir}$, $k^A_r$ to
obey  \begin{eqnarray} d_{I\,r(u}d^I_{vs)} &=& 0 \;,\nonumber\\ [.4ex] {} \left(d^J_{r(u}\, d^I_{v)s}
-d^J_{uv}\,d^I_{rs} + d_{K\,rs} d^K_{uv}\, \eta^{IJ}\right)g_J^s &=& f_{r(u}{}^s d^I_{v)s}
\;,\nonumber\\[.4ex] 3f_{[pq}{}^u f_{r]u}{}^s -g^s_I\, d^{I}_{u[p} f_{qr]}{}^u &=& 0 \;,\nonumber\\
[.4ex] {} X_{rs}{}^t\equiv d^I_{rs}\,g_I^t-f_{rs}{}^t&=&-k_r^A c_{A\,s}^t\nonumber\\ [.4ex] {}X_{r\,IJ}
\equiv 4 g_{[I}^sd_{J]\,rs}   &=& 2\,k_r^A\,c_{A\,IJ}\nonumber\\ [.4ex] {} f_{rs}{}^t g_I^r -
d^J_{rs}\, g_J^t g_I^r &=& 0 \;,\nonumber\\[.4ex] g_K^r g_{[I}^{s}d^{\vphantom{s}}_{J]sr} &=& 0
\;,\nonumber\\[.4ex] g_I^r g^{Is} &=& 0 \;, \nonumber\\[.4ex] k^A_r g^{Ir} &=& 0\,. \label{conaction}
\end{eqnarray} Notice also the relations which are valued at least when contracted with  $k^A_r$ matrix
\begin{eqnarray} {} X_{sA}{}^B&=& c_A{}^t_s k^B_t \;,\nonumber\\ [.4ex] {}  c_A{}^t{}_{(s|}  d_{I\,
|r)t}   &=& c_{A\, IJ} d^J_{rs}\;,\nonumber\\[.4ex] {} c_A{}^t{}_s g_{I}^s  &=& -2 c_{A\, IJ} g^{Jt} \;.
\end{eqnarray}

\section{Some useful identities for 6d gamma matrices} \setcounter{equation}{0}

We use the metric of mostly minus  signature $\eta^{ab}= diag (+,-,-,-,-,-)$. The $4\times 4$
matrices $\gamma^a_{\alpha\beta}$ and $\tilde{\gamma}^{a\gamma\delta}$ obey
\begin{eqnarray}
 \label{6Dgammas+}
\gamma^a_{\alpha\beta}=-\gamma^a_{\beta\alpha}= {1\over
2}\epsilon_{\alpha\beta\gamma\delta}\tilde{\gamma}^{a\gamma\delta} \; , \qquad
(\gamma^{(a}\tilde{\gamma}{}^{b )})_{\alpha}{}^{\beta} = \eta^{ab}\delta_{\alpha}{}^{\beta}\; \; ,
\qquad \nonumber \\  \gamma^a_{\alpha\beta}\tilde{\gamma}_a^{\gamma\delta}= -4
\delta_{[\alpha}{}^{\gamma} \delta_{\beta ]}{}^{\delta} \; , \qquad \gamma^a_{\alpha\beta} {\gamma}_{a
\gamma\delta}= -2 \epsilon_{\alpha\beta\gamma\delta} \; , \qquad  \\ tr(\gamma_a\tilde{\gamma}^b)=
4\delta_a^b\; , \qquad tr(\gamma_{cd}{\gamma}^{ab})= -8\delta_{[c}^a\delta_{d]}^b\; , \qquad  \nonumber
\\ tr(\gamma^{abc}\tilde{\gamma}_{def})= -24\delta_{[d}^a\delta_{e}^b\delta_{f]}^c - 4
\epsilon^{abc}{}_{def}\; , \qquad   \\ \delta_{\alpha}^{[\gamma}\delta_{\beta}^{\delta ]}= -{1\over 4}
\gamma^{a}_{\alpha\beta} \tilde{\gamma}{}_{a}^{\gamma\delta} \; , \qquad
\delta_{\alpha}^{(\gamma}\delta_{\beta}^{\delta )}= -{1\over 48} \gamma^{abc}_{\alpha\beta}
\tilde{\gamma}{}_{abc}^{\gamma\delta}\; , \qquad \nonumber \\
\gamma^{ab}{}_{\alpha}{}^{\beta}{\gamma}_{ab}{}_{\gamma}{}^{\delta}= -8 \delta_{\alpha}{}^{\delta}
\delta_{\gamma}{}^{\beta}+2 \delta_{\alpha}{}^{\beta} \delta_{\gamma}{}^{\delta}  \; , \qquad     \\
\gamma^{abcdef}{}_{\alpha}{}^{\beta}= - \epsilon^{abcdef}\delta_{\alpha}{}^{\beta}  \; , \qquad
\nonumber \\ \gamma^{abcde}{}_{\alpha\beta}= - \epsilon^{abcdef}\gamma_{f\;\alpha\beta} \; , \qquad
\tilde{\gamma}^{abcde \alpha\beta}= \epsilon^{abcdef} \tilde{\gamma}_f^{\alpha\beta}
 \; , \qquad   \nonumber \\
 \gamma^{abcd}{}_{\alpha\beta}=
 {1\over 2}\epsilon^{abcdef}\gamma_{ef\; \alpha\beta} \; , \qquad
 \tilde{\gamma}^{abcd \alpha\beta}=  - {1\over 2}\epsilon^{abcdef}\tilde{\gamma}_{ef}^{\alpha\beta}
\; . \quad \nonumber \\ \label{6Dg3=A} \gamma^{abc}{}_{\alpha\beta}=\gamma^{abc}{}_{(\alpha\beta )}= +
{1\over 3!}\epsilon^{abcdef}\gamma_{def\;\alpha\beta} \; , \qquad  \nonumber \\ \tilde{\gamma}^{abc
\alpha\beta}=\tilde{\gamma}^{abc (\alpha\beta )}=  - {1\over
3!}\epsilon^{abcdef}\tilde{\gamma}_{def}^{\alpha\beta} \;  \qquad \end{eqnarray} \begin{eqnarray}
\label{gagabc=} \gamma^{a}{}_{\alpha\beta} \gamma_{abc\, \gamma\delta} =
2\epsilon_{\alpha\beta\gamma\kappa}  \gamma_{bc\, \delta}{}^{\kappa} -2 \gamma_{[b|\, \alpha\beta}
\gamma_{|c]\, \gamma\delta} \; , \nonumber  \qquad \\ \gamma^{a}{}_{\alpha\beta} \gamma_{abcd\,
\gamma}{}^{\delta} = 2\epsilon_{\alpha\beta\gamma\kappa}  \tilde{\gamma}_{bcd}^{\kappa\delta} -3
\gamma_{[b|\, \alpha\beta} \gamma_{|cd]\, \gamma}{}^{\delta} \; . \qquad \end{eqnarray}

\section{On identities for identities and dependence of Bianchi identities of the tensorial hierarchy in D=6 (1,0) superspace }

Let us discuss  the interrelations between different components of the Bianchi identities ({\it BIs}). In supergravity  the most known of such interdependencies  is described by the Dragon theorem \cite{Dragon:1978nf}. However,   a dependence of higher dimensional components of the differential form  {\it BIs} on the lower dimensional ones is more universal. A convenient  tool for  establishing Dragon-like theorems is provided by the so--called identities for identities ({\it Ids for Ids})
\cite{Sohnius:1980iw} which were used intensively already in 80th (see e.g. \cite{Howe:1981tp}). In our case these are the consistency conditions
\begin{eqnarray}
 \label{J4=IdfId3}
 && J_4^r:= DI_3^r+ g_I^r I_4^I =0\; , \qquad \\
  \label{J5=IdfId4}
 && J_5^I:= DI_4^I+2
 d_{st}^I {\cal F}^s\wedge I_3^t+ g^{Ir} I_{5r} =0 \; , \qquad \\
 \label{J6=IdfId5}
 && J_{6r}:= DI_{5r} - 2 I_4^I \wedge {\cal F}^sd_{Isr}- 2{\cal H}_3^I \wedge I_3^sd_{Isr}+ k_r^{A}I_{6A}=0\; , \qquad \\
 \label{J7=IdfId6}
 && k_r^AJ_{7A} := k_r^ADI_{6A} + k_r^Ac_{A \, s}^{u}\,   I_3^s \wedge  {\cal H}_{4 u} - 2 k_r^Ac_{AIJ}  {\cal H}_3^I \wedge I_4^J + k_r^Ac_{A \, s}^{u}\,   {\cal F}^s \wedge I_{5 u}  =0
 \;  \qquad
\end{eqnarray}
for the  {\it BIs}  (\ref{DcF=}), (\ref{DcHI=}), (\ref{DcH4=}) and (\ref{DcH5=}) which we have denoted by  \begin{eqnarray}
 \label{IqLq=}
 I_q^{\Lambda_q}:=(I_3^r, I_4^I, I_{5r}, k_r^{A}I_{6A})= 0
  \; . \end{eqnarray}

The cancelations of the contributions which are not proportional to $I_q^{\Lambda_q}$ in the {\it r.h.s.}s of (\ref{J4=IdfId3})--(\ref{J7=IdfId6}) occur due to the properties (\ref{conaction}) of the constant tensor. Actually the relations (\ref{conaction}) can be obtained by requiring this cancelation.

With our constraints a number of lower dimensional components of the {\it BIs}  are satisfied due to the algebraic structure of the superfield strengths. Omitting these, we find the following decomposition of the superspace Bianchi identities
\begin{eqnarray}
 \label{I3=EEE}
I_3^r &=& {1\over 2}E^b \wedge E^{\alpha i}\wedge E^{\beta j} I_{\alpha i\,  \beta j\, b}^r+ {1\over 2}E^c \wedge E^b \wedge E^{\alpha i} I_{\alpha i\,  bc}^r    + {1\over 3!}E^c \wedge E^b \wedge E^{a} I_{a   bc}^r  \; , \qquad \\
  \label{I4=EEE}
 I_4^I&=&
 {1\over 3!} E^c \wedge E^{\alpha i}  \wedge E^{\beta j} \wedge E^{\gamma k}
 I_{\alpha i\, \beta j\, \gamma k\, c}^I+
  {1\over 4}E^c \wedge E^b \wedge E^{\alpha i}\wedge E^{\beta j} I_{\alpha i\, \beta j\, bc}^I+ \qquad \nonumber  \\ && + {1\over 3!}E^c \wedge E^b \wedge E^a \wedge E^{\alpha i} I_{\alpha i\,  abc}^I    + {1\over 4!}E^d \wedge E^c \wedge E^b \wedge E^{a} I_{a   bcd}^I \; , \qquad \\
 \label{I5=EEE}
 I_{5r}&=&
 {1\over 2\cdot 3!} E^c \wedge E^b \wedge E^a \wedge E^{\alpha i}  \wedge E^{\beta j}
 I_{\alpha i\, \beta j\, abc\, r}+
  {1\over 4!}E^d \wedge E^c \wedge  E^b \wedge  E^a  \wedge E^{\alpha i} I_{\alpha i\,  abcd\, r}+ \qquad \nonumber  \\ && + {1\over 5!}  E^e \wedge E^d \wedge E^c \wedge E^b \wedge E^{a} I_{abcde\, r}\; , \qquad \\
 \label{kI6=EEE}
 k_r^A &I_{6A} &= {1\over 2\cdot 4!}  E^d  \wedge E^c \wedge E^b \wedge E^a \wedge E^{\alpha i}  \wedge E^{\beta j}
 k_r^A I_{\alpha i\, \beta j\, abcd\, A} + \qquad \nonumber  \\ && +  {1\over 5!}E^d \wedge E^c \wedge  E^b \wedge  E^a  \wedge E^{\alpha i} k_r^A I_{\alpha i\,  abcde\, A} + \qquad \nonumber  \\ && + {1\over 6!} E^f \wedge E^e \wedge E^d \wedge E^c \wedge E^b \wedge E^{a} k_r^A I_{abcdef\, A}\;
 \; . \qquad
\end{eqnarray}
Generically, the first nontrivial component in the $q$-form {\it BI} $I_q^{\Lambda_q}$ has dimension $q-1$, i.e. carries all--but--two bosonic indices. The only exception is the nontrivial  identity $  I_{\alpha i\, \beta j\, \gamma k\, c}^I=0$ in (\ref{I4=EEE}) which results in identification $\Psi_i^I= -i/2D_{\alpha i}\Phi$.

Substituting (\ref{I3=EEE})--(\ref{kI6=EEE}) into the {\it Ids for Ids}
(\ref{J4=IdfId3})--(\ref{J7=IdfId6}) one can study the interdependence of different components of different {\it BIs}  and establish some Dragon--like theorems for the tensorial hierarchies in 6D superspace. We will not perform here such a complete study, but just mention a few particular results.

\begin{itemize}

\item
For instance,  one can establish that the $SU(2)$ tensorial part of the dim 4 component of the {\it BI}  (\ref{DcH4=}) for the 4-form superfield strength, $I_{\alpha (i\, \beta j)\, abc \; r}=0$, is dependent on the lower dimensional {\it BI}s, and that the independent parts of the SU(2) singlet  $\epsilon^{ij}I_{\alpha i\, \beta j\, abc \; r}=0$, are the duality equation (\ref{cH4=Fpd}), ${\cal I}_{abcd\; r}\equiv {\cal I}_{[abcd]\; r} =0$, and the auxiliary superfield equation (\ref{Yphd=ff}),
${\cal I}_{ij\; r}\equiv {\cal I}_{(ij)\; r}= 0$:  \begin{eqnarray}
 \label{I4r4=I+I}
 I_{\alpha i\, \beta j\, abc \; r}=- 2i \epsilon_{ij} \gamma^d_{\alpha \beta} {\cal I}_{abcd \; r}-2i  \epsilon_{ij} \gamma_{abc\, \alpha \beta} {\cal I}_{ij\; r }
\;  . \qquad
\end{eqnarray}

\item
The independent parts of next-to-higher dimensional components of the {\it BIs}, the ones with all--but--one  bosonic indices, are defined by nontrivial solutions of the equation
\begin{eqnarray}
 \label{I=dim-q-3}
&& \gamma^a_{\alpha\beta} \epsilon_{ij}\tilde{I}_{\gamma k\; a, b_1...b_{q-2}}{}^{\Lambda_q}  + cyclic (\alpha i\, , \beta j\, \gamma k) =0 \; , \qquad \end{eqnarray}
where
\begin{eqnarray}
\label{tI=I+}
 && \tilde{I}_{\gamma k\; a, b_1...b_{q-2}}^{\Lambda_q} = \left( I_{\gamma k\; ab}^r, I_{\gamma k\; ab_1b_2}^I, \tilde{I}_{\gamma k\; a,b_1b_2b_3\; r}
, \tilde{I}_{\gamma k\; a,b_1b_2b_3b_4\; A}k_r^A
\right)\; , \\ \label{tI5=I+}
&& \tilde{I}_{\gamma k\; a, b_1b_2b_3\; r}= I_{\gamma k\; ab_1b_2b_3\; r} + \propto \eta_{a[b_1} \Phi^I I_{\gamma k\; |b_2b_3]}^sd_{Isr}\; , \qquad
\\ \label{tI6=I+} && \tilde{I}_{\gamma k\; a, b_1b_2b_3b_4\; A}k_r^A=  I_{\gamma k\; ab_1b_2b_3b_4\; A}k_r^A +  \propto   k_r^Ac_{AIJ}\Phi^I \eta_{a[b_1}  I_{\gamma k\; |b_2b_3b_4]}^J
 \; . \qquad
\end{eqnarray}
It is easy to see that Eq. (\ref{I=dim-q-3}) is solved by $\tilde{I}_{\gamma k\; a, b_1...b_{q-2}}{}^{\Lambda_q}\propto  {\gamma}_{a\gamma\delta} {\cal I}^\delta_{k\; b_1...b_{q-2}}{}^{\Lambda_q}$.  Hence the independent part of the  next-to-higher dimensional components of the {\it BIs} are given by their gamma--traces,
\begin{eqnarray}
 \label{Ifbbb=}
 {\cal I}^\beta_{k\; b_1...b_{q-2}}{}^{\Lambda_q}= \tilde{\gamma}^{a\beta\gamma} {I}_{\gamma k\; a, b_1...b_{q-2}}{}^{\Lambda_q}=0\; .  \qquad
\end{eqnarray}

\item It is easy to check that the highest dimensional components of the differential form {\it BIs}, the ones with all  bosonic indices, ${I}_{b_1...b_{q}}{}^{\Lambda_q} = \left( I_{b_1...b_{3}}^r, I_{b_1...b_{4}}^I, {I}_{b_1...b_{5}\; r}
, {I}_{b_1...b_{6}\; A}k_r^A
\right)$ are dependent. They are satisfied identically due to the lower dimensional  {\it BIs} (and their derivatives) and thus do not require a separate study of their consequences.
\end{itemize}

\end{appendix}

\end{document}